\documentclass[12pt]{iopart}

\usepackage{epsfig}

\usepackage{amssymb}

\begin{document}

\title{Landau theory for magnetic and structural transitions in CeCo$_{0.85}$Fe$_{0.15}$Si}

\author{William Gabriel Carreras Oropesa$^1$, V\'ictor F. Correa$^{1,2}$, Juli\'an G. Sereni$^{1,2}$, Daniel J. Garc\'{\i}a$^{1,2}$, Pablo S. Cornaglia$^{1,2}$}

\address{$^1$ Centro At{\'o}mico Bariloche and Instituto Balseiro, CNEA, 8400 Bariloche, Argentina}
\address{$^2$ Consejo Nacional de Investigaciones Cient\'{\i}ficas y T\'ecnicas (CONICET), Argentina}

\begin{abstract}
We present a phenomenological analysis of the magnetoelastic properties of CeCo$_{0.85}$Fe$_{0.15}$Si at temperatures close to the N\'eel transition temperature $T_N$. Using a Landau functional we provide a qualitative description of the thermal expansion, magnetostriction, magnetization and specific heat data. We show that the available experimental results [Journal of Physics: Condensed Matter {\bf 28} 346003 (2016)] are consistent with the presence of a structural transition at  $T_s\gtrsim T_N$ and a strong magnetoelastic coupling. The magnetoelastic coupling presents a Janus-faced effect: while the structural transition is shifted to higher temperatures as the magnetic field is increased, the resulting striction at low temperatures decreases.
The strong magnetoelastic coupling and the proximity of the structural transition to the onset temperature for magnetic fluctuations, suggest that the transition could be an analogue of the tetragonal to orthorhombic observed in Fe-based pcnictides. 
\end{abstract}
\noindent{\it Keywords: Magnetism, Magnetostriction, Landau functional, Phase transitions \/}

\section{Introduction}
Ce based compounds have attracted considerable attention over the years due to their wide range of physical properties which include unconventional superconductivity~\cite{steglich2016foundations}, heavy fermion behavior~\cite{stewart1984heavy}, magnetism, non-Fermi liquid behavior and quantum phase transitions~\cite{lohneysen2007fermi}. 
In these compounds the properties depend strongly on the hybridization of the 4f Ce$^{3+}$ orbital to the conduction band and on the dimensionality. 
The crystalline environment of the Ce$^{3+}$ ions determines the degree of localization of the 4f electrons and the magnetic interactions between them. As a result, these systems can present magnetic ground states with ordered local magnetic moments or heavy fermion behavior where the magnetic moments are Kondo screened.  External pressure or chemical doping may induce a transition between these phases. The role of the dimensionality manifests itself in, e.g., the layered 115 compounds, CeMIn$_5$ (M=Rh,Co, Ir) where decreasing the coupling between layers leads to an increase in the superconducting transition temperature~\cite{Petrovic2001,Movshovich2001,WALKER1997303,Facio2015}.
These compounds share a number of common features with the cuprate superconductors that have made them a proxy in the quest to understand high temperature superconductivity \cite{0295-5075-53-3-354}. 

The rich variety of behavior presented by Ce-based compounds seems to be, in general, dominated by electron-electron correlations. In Ce mono-pnictides, however, strong signatures of the coupling between the magnetic and elastic degrees of freedom have been reported~\cite{hulliger1979rare,takeuchi1998magnetoelastic,rossat1980specific,siemann1979enhanced}.
More recently, in the CeCo$_{0.85}$Fe$_{0.15}$Si compound, a strong signature in the thermal expansion ($\Delta L/L\sim 10^{-4}$) was observed at the magnetic transition~\cite{Correa2016}, indicating the presence of a significant magnetoelastic coupling.  

\begin{figure}[t]
\includegraphics[width=\columnwidth]{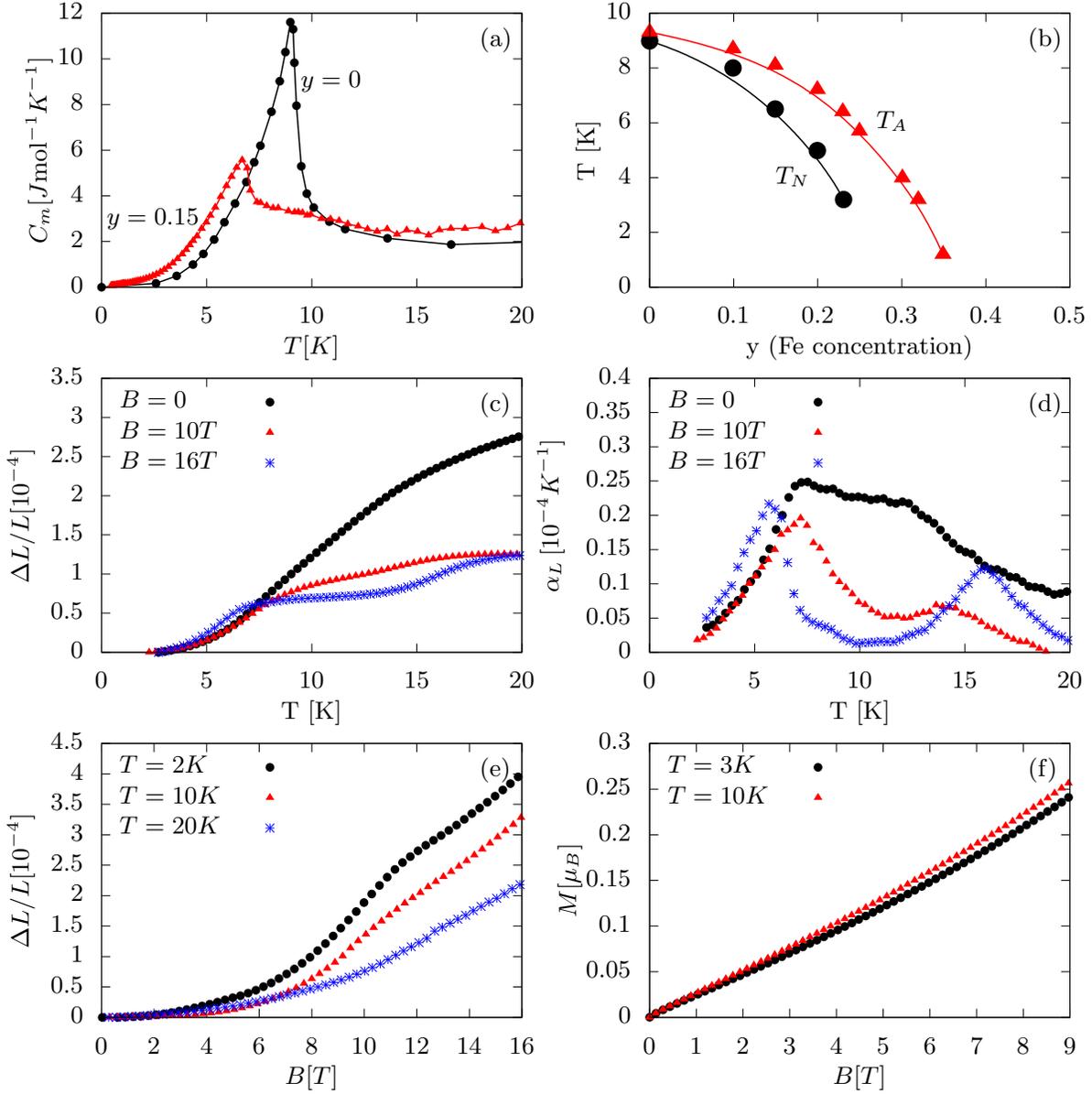}
	\caption{ (a) Magnetic contribution to the specific heat $C_m$ for CeCoSi and for CeCo$_{0.85}$Fe$_{0.15}$Si at $B=0$. (b) Evolution of the N\'eel temperature  $T_N$ and the specific heat anomaly temperature $T_A$ as a function of the Fe concentration $y$ in CeCo$_{1-y}$Fe$_y$Si compounds. The lines are a guide to the eye. Data taken from \cite{Sereni2014}. (c) Linear thermal expansion $\Delta L/L$ as a function of the temperature for different external magnetic fields. (d) Linear thermal expansion coefficient $\alpha_L$. (e) Magnetostriction at different tempertures. (f) Uniform magnetization as a function of the external magnetic field, for different temperatures.}
\label{fig:experiment}
\end{figure}

CeCo$_{1-y}$Fe$_y$Si compounds range from CeCoSi, which presents a second order transition to an antiferromagnetic state at $T_N=8.8~K$, to CeFeSi which is a paramagnetic Fermi liquid. 
As the concentration of Fe ($y$) increases, the N\'eel temperature $T_N$, as deduced from the peak in the specific heat at the transition, decreases and the peak becomes weaker [see figure \ref{fig:experiment}(a)]. The behavior of the specific heat \cite{Sereni2014} suggests a chemical pressure effect due to the substitution of Co by Fe. This leads to a suppression of the N\'eel transition and the development of a bump in a way that resembles a dimensional crossover from 3D to 2D magnetism \cite{Sengupta2003}. 
The antiferromagnetism is completely suppressed [see figure \ref{fig:experiment}(b)] for $y\gtrsim 0.23$~\cite{Sereni2014}. Interestingly, an anomaly in the specific heat at a temperature $T_A>T_N$ was identified in \cite{Sereni2014} which was interpreted as an onset of large magnetic fluctuations in the paramagnetic phase near the N\'eel transition.

In the $y=0.15$ compound a strong dependence of the linear expansion on the magnetic field was also observed [see figure \ref{fig:experiment}(c)]. A peak in the thermal expansion, which was interpreted as stemming from a structural transition, is obtained at a temperature $T\sim T_s$ and another at the N\'eel transition [see figure \ref{fig:experiment}(d)]. Both peaks show a strong magnetic field dependence, while the peak at $T_N$ shifts to lower temperatures as the magnetic field is increased, the peak at $\sim T_s$ shifts to higher temperatures. Figure \ref{fig:experiment} summarizes the main experimental observations for the $y=0.15$ Fe doping concentration~\cite{Correa2016}.
The magnetostriction presents perhaps the more puzzling behavior  [see figure \ref{fig:experiment}(e)]. The shift to higher temperatures of the structural transition with increasing magnetic field would seem to imply an enhanced structural distortion at low temperatures as the magnetic field is increased. The experimental results present precisely the opposite behavior at low temperatures.

Motivated by these recent experiments on the magnetoelastic properties of polycrystalline CeCo$_{0.85}$Fe$_{0.15}$Si  \cite{Correa2016} that show a strong magnetostructural coupling and suggest the presence of a structural transition, we analyze the possibility of the presence of such structural transition to explain the observed experimental data. To that aim we propose a Landau free energy to describe a magnetic transition and a structural transition including a magnetoelastic coupling. The qualitative agreement obtained with the available experimental data indicates that the latter is consistent with the presence of a structural transition at a temperature $T_s\sim T_A\gtrsim T_N$.

\section{Landau theory}
For the doping $y=0.15$ the material shows a clear N\'eel transition at $T_N \simeq 6.5~K$. The anomaly in the specific heat evolves continuously from a textbook transition in the $y=0$ compound until it vanishes at a doping $y\sim 0.23$. The linear thermal expansion coefficient $\alpha_L$ presents a wide peak ranging from $T\sim T_N$ to $T\sim 13 K$ that in the presence of an external magnetic field splits into two peaks [see figure \ref{fig:experiment}(d)]. 

To describe the magnetic transition under an external magnetic field we propose a Landau functional in terms of the staggered magnetization $m=m_A-m_B$ and a uniform magnetization $M=m_A+m_B$, where $m_A$ and $m_B$ correspond to two sublattices which are coupled antiferromagnetically. We also include an order parameter $\delta$ to describe a structural transition. The structural transition produces a lattice striction that for simplicity we assume proportional to $\delta$ ($\Delta L/L=\delta$). As usual we consider that it is possible to make a series expansion of the free energy in terms of the order parameters close to the transitions and consider the lowest order terms allowed by symmetry. 
The free energy in units of $E_0=250J$mol$^{-1}$ can be written as
\begin{equation}\label{eq:freeen}
    \Phi= \Phi_m + \Phi_h+ \Phi_x +\Phi_{xm},
\end{equation}
where the magnetic transition is described by 
\begin{equation}
\Phi_m = -a_m\left(1-\frac{T}{T_N}\right)m^2 + b_m m^4 + c_1 (m M)^2 + \frac{1}{2\chi_U} M^2.
\end{equation}
Here the first two terms, where $a_m>0$ and $b_m>0$, correspond to the standard functional to describe a mean field second order transition. The third term with $c_1>0$ is the competition between the staggered and uniform magnetizations and the last term is the energy associated with an uniform magnetization.
\begin{equation}
\Phi_h = k_1 m^2 h^2 -M h,
\end{equation}
with $k_1>0$, describes the lowest order coupling terms of the magnetic field to the magnetizations\footnote{The coupling term between the staggered magnetization and the magnetic field is expected to depend on the angle $\theta$ between them as $\cos^2\theta$ (see e.g. \cite{khomskii2010basic}). For a polycrystalline sample we consider here, for simplicity, this coupling term as the result of an average over $\theta$. }.

The structural transition is described by 
\begin{equation}
    \Phi_x = -a_x\left(1-\frac{T}{T_x}\right)\delta^2 +c_{x}\delta^3+ b_x \delta^4 
\end{equation}
where $a_x>0$, $b_x>0$ and a finite $c_x$ sets the sign of the deformation $\delta$, $c_x>0$ corresponding to a contraction ($\delta<0$) below the transition temperature. For finite $0<c_x\ll b_x$ the transition is weak first order with a jump in the order parameter $\sim c_x/b_x$.
Finally, for the magnetoelastic coupling we expect terms of the form $\gamma_{n}^{\pm} (M^2\pm m^2)\delta^n$, for $n=1,2,\ldots$, where the $+$ sign corresponding to a local coupling $\propto (m_A^2+m_B^2)$, and the $-$ sign to a non-local coupling $\propto m_Am_B$. As we will describe below, $n=2$ terms with $\gamma_2^{+}+\gamma_2^{-}<0$ are crucial to describe the shift to higher temperatures of the structural transition as the magnetic field is increased, while the $n=4$ term with $\gamma_4^{+}\sim\gamma_4^{-}$ allows to explain the magnetostriction results at low temperature. The minimal magnetoelastic coupling terms that allow to describe qualitatively the available experimental data read
\begin{equation}
	\Phi_{xm} = \gamma_2 (m^2+ M^2)\delta^2 + \gamma_{4M} M^2\delta^4 + \gamma_{4m} m^2\delta^4,
\end{equation}
where $\gamma_{4m}\equiv\gamma^+_4-\gamma^-_4\ll \gamma_{4M}\equiv\gamma^+_4+\gamma^-_4$ and we have, for simplicity, assumed $\gamma_2^-=0$.
\section{Determination of the Landau free energy parameters}

The structural transition temperature $T_s\sim 12.5 K$ is determined by the high temperature peak in the linear thermal expansion ($\alpha_L$) in the absence of an external magnetic field. Guided by the behavior of $\alpha_L(T)$ as the magnetic field is increased, we assume that the broad peak observed in $\alpha_L(T)$ for $B=0$ is composed by two peaks, one at $T_N$ and the other at $T_s$.

For small external fields ($h\to 0$) and high temperatures ($T>T_s>T_N$) we have $m=0$, $\delta=0$ and $M\to 0$ and the magnetization $M$ is simply given by $M=\chi_U h$.
The experimental data shows an approximately linear behavior of $M$ at low fields and an constant $\chi_U$ in the temperature range where the transitions take place. We measure the magnetization per atom $M$ in terms of its saturation value $g_J\mu_B J$ and set $\chi_U=1/80$ Tesla$^{-1}$ which is consistent with the saturation field obtained extrapolating the experimental data\footnote{Hund's rules applied to Ce$^{3+}$'s $4f$ electron result in $J=5/2$ and a Land\'e factor $g_J=6/7$. The lowest lying multiplet is however a doublet due to the presence of a crystal field.}.

Assuming a weak effect of the structural order on the uniform magnetization $M$, the structural transition temperature reads (for $c_x^2\ll a_x b_x$) 
\begin{eqnarray}
	T_s(h)&=\left(1-\frac{\gamma_2M^2}{a_x} +\frac{9c_x^2}{32 a_x(b_x+\gamma_{4M}M^2)} \right)T_x \nonumber \\&\simeq  \left(1-\frac{\gamma_2\chi_U^2 h^2}{a_x}\right)T_x
\end{eqnarray}  
which for $\gamma_2<0$ leads to an increase of the transition temperature with increasing magnetic field. The observed positive shift of $3K$ in $T_s$ for $h=16T$ [see figure \ref{fig:experiment}(d)] indicates $\gamma_2\sim -9a_x$. 
For temperatures larger than the magnetic transition temperature, the structural order parameter can be described by a functional
\begin{equation} \label{eq:elast}
	\Phi(T>T_N)  \simeq -a_x(h)\left(1-\frac{T}{T_s(h)}\right)\delta^2 +c_{x}\delta^3+ b_x(h) \delta^4 
\end{equation}
where
\begin{eqnarray}
    a_x(h)=a_x-\gamma_2 \chi_U^2 h^2,\\
	b_x(h)=b_x+\gamma_{4M} \chi_U^2 h^2,
\end{eqnarray}
and the mean field solution for the order parameter is
\begin{equation}\label{eq:delta}
	\delta(T_N<T<T_s)\simeq -\frac{3}{8}\frac{c_x}{b_x(h)}-\sqrt{\frac{a_x}{2 b_x(h)T_x}\left( T_s(h)-T \right)}.
\end{equation}

For $T_N<T<T_s$ we have $M\simeq\chi_U^\star h$ where $\chi_U^\star=\chi_U-2\gamma_2 \delta^2-2\gamma_4 \delta^4$ is the effective magnetic susceptibility. For $T\to T_s(h)$, $\delta^2\ll 1$ and there is an increase of the susceptibility as $\gamma_2<0$. i.e., the magnetization increases as the temperature decreases below $T_x(h)$ in a fixed external magnetic field.

An external magnetic field and the presence of a distortion shifts the magnetic transition to lower temperatures:
\begin{equation}
	T_N(h)\sim \left(1-\frac{h^2 (c_1{\chi_U^\star}^2 +k_1)+\gamma_2\delta^2(T_N)}{a_m}\right)T_N
\end{equation}
The observed reduction of $T_N$ of $\sim 1.5K$ at $h=16T$ imposes the constraint $a_m\sim c_1/5+1280 k_1$ on the functional parameters.

At the N\'eel transition, the magnetoelastic coupling produces a kink in $\delta$ due to the onset of $m$. For $T\lesssim T_N$ and $h=0$ we have
\begin{equation}
	\delta(T\lesssim T_N)\sim -\sqrt{\frac{a_m b_x \left(1-\frac{T}{T_N}\right)}{2 b_m b_x-2 \gamma _2^2}+\frac{ \gamma _2 a_x \left(1-\frac{T}{T_x}\right)}{2 \gamma _2^2-2 b_m b_x}},
\end{equation}
where we considered the lowest order tems in the coupling and dropped terms of order $c_x/b_x$.

Since there is no signature in the specific heat of the structural transition, the jump in the specific heat at the magnetic transition $\Delta C^m\sim a_m^2T_N/2 b_m$ must be much higher that the corresponding one at the structural $\Delta C^x\sim a_x^2T_x/2 b_x$ transition. We also require the latent heat at the structural transition $\Delta Q_x \sim \frac{9 a_x c_x^2}{64  b_x^2}$ to be small $\Delta Q_x\ll T_N \Delta C^m$. This sets the constrains $a_x^2/b_x\ll a_m^2/b_m$ and $c_x\ll a_m b_x/\sqrt{a_x b_m}$ on the parameters. Additionally, we set the parameters to satisfy $m(T\to 0)\sim \sqrt{{a_m}/{2b_m}}\sim 1$  and $\delta(T\to 0)\sim \sqrt{{a_x}/{2b_x}}\sim 0.0005$.

Table \ref{tab:params} lists the parameters used to obtain a qualitative description of the experimental data.

\begin{center}
\begin{table}
\caption{Landau functional parameters}	\label{tab:params}
\begin{tabular}{|l|c|c|}
\hline
	Parameter&&value\\
	\hline

	Global energy scale&$E_0$&$250\, J$ mol$^{-1}$\\
	Uniform susceptibility &$\chi_U$& $1/80$ Tesla$^{-1}$ \\
	Temperature scale (magnetic) &$T_N$& 7.5 $K$ \\
	Temperature scale (structural)& $T_x$& 12.5 $K$ \\
	\hline
	\hline
	Functional term& &value\\ 
	\hline
	-$m^2(1-T/T_N)$	& $a_m$&0.5\\
$m^4$	&$b_m$& 0.35\\
$m^2 M^2$	&$c_1$& $0.625$\\
$m^2 h^2$	&$k_1$& $2.93\times 10^{-4}$\\
-$\delta^2(1-T/T_x)$	&$a_x$&3000\\
$\delta^3$	&$c_x$& 10000\\
$\delta^4$	&$b_x$& $7.5 \times 10^9$\\
	$(m^2+M^2)\delta^2$	&$\gamma_{2}$ & $-27000$\\
$M^2\delta^4$	&$\gamma_{4M}$& $1.875 \times 10^{12}$\\
$m^2\delta^4$	&$\gamma_{4m}$& $3.75 \times 10^{10}$\\
\hline
\end{tabular}
\end{table}
\end{center}
\section{Numerical results}
\begin{figure}[th]
\includegraphics[width=\columnwidth]{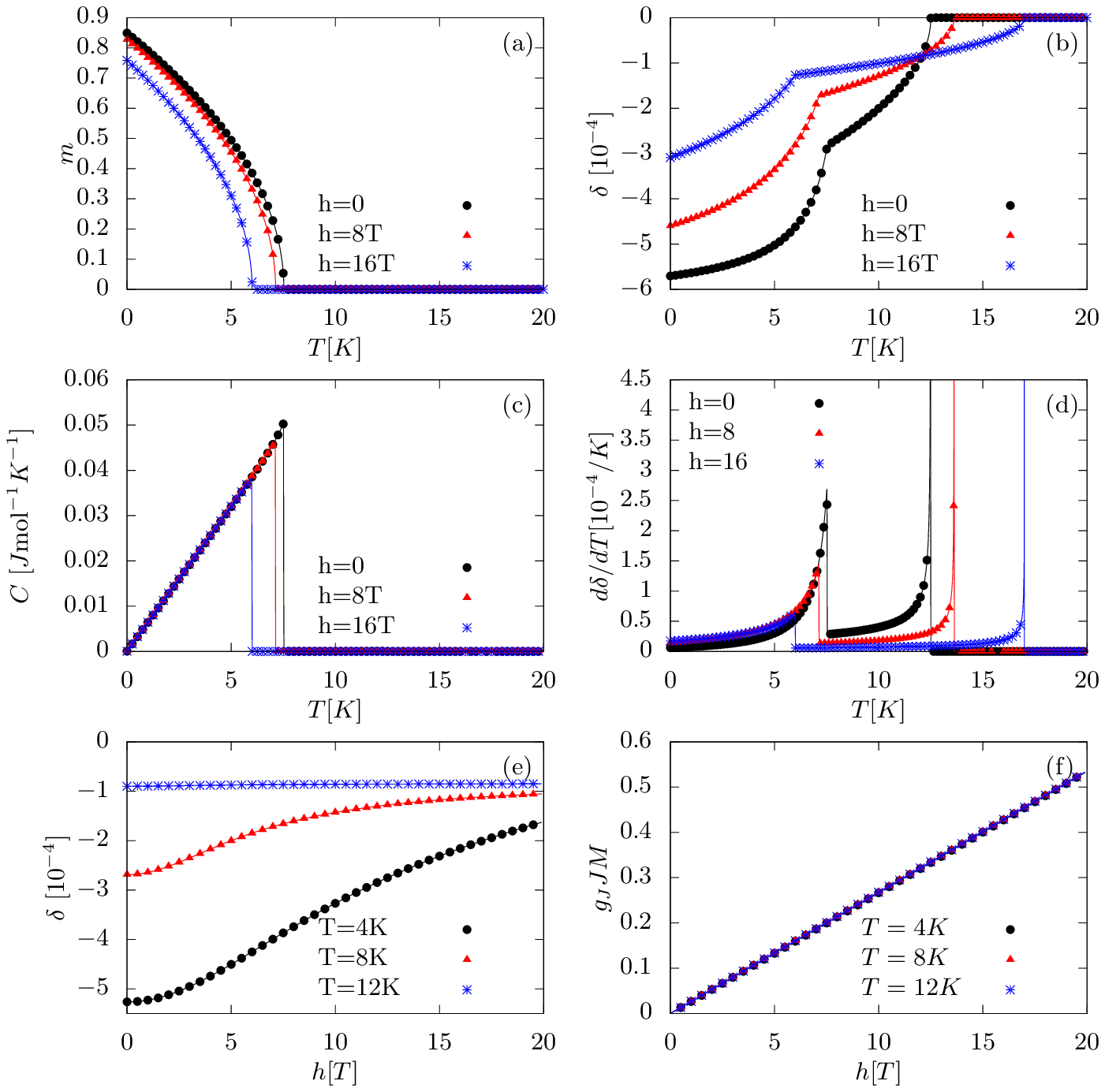}
	\caption{(a) Staggered magnetization $m$, (b) structural order parameter $\delta$, (c) specific heat $C$, and (d) linear distortion parameter $d\delta/dT$, as a function of the temperature for different external magnetic fields applied. (e) Structural order parameter and (f) uniform magnetization $M$, a function of the external magnetic field. }
\label{fig:theory}
\end{figure}
Figures \ref{fig:theory}(a) and \ref{fig:theory}(c) present the staggered magnetization $m$ and specific heat $C$ data.  The antiferromagnetic transition is shifted towards lower temperatures when the magnetic field is increased as it can be seen in the staggered magnetization $m$ and specific heat $C$ data. The structural order parameter $\delta$ is shown in figure \ref{fig:theory}(b) as a function of the temperature for different values of the external magnetic field. As the magnetic field increases, the structural transition temperature (where $\delta$ acquires a nonzero value) increases. Note, however, that as a consequence of the $M^2\delta^4$ term in the free energy, the rate of increase of $|\delta|$  decreases as $h$ increases. This term in therefore necessary to explain the behavior of magnetostriction at low temperatures [see figures \ref{fig:experiment}(e) and \ref{fig:theory}(e)].
The magnetoelastic coupling $\propto \delta^4$ hardens the lattice as the magnetization increases. The qualitative behavior of the experimental data can be accounted including a much larger coupling to the squared magnetization $M^2$ than to the squared staggered magnetization $m^2$. As mentioned above this indicates a non-local magnetoelastic coupling $\propto m_A m_B \delta^4$.

At the N\'eel transition, $\delta$ presents a kink and a faster absolute value increase for decreasing temperature, as a consequence of the magnetoelastic coupling $\gamma_2 m^2 \delta^2$ (with $\gamma_2<0$) and the increase of $m^2$ for $T<T_N$. The mean-field solutions for $d\delta/d T$ present a divergent behavior $\propto (T_s-T)^{-1/2}$  [see (\ref{eq:delta})] at the structural transition and a discontinuity at the N\'eel transition [see figure \ref{fig:theory}(d)]. The experimental results, however, are obtained for polycrystalline samples where a distribution of transition temperatures is expected. To take this into account in an approximate way we performed a Gaussian convolution of the structural order parameter $\tilde{\delta}(T)=(\delta * G)(T)$ which is a convolution of $\delta(T)$ with a Gaussian function\footnote{We are assuming here the same distribution for the magnetic and the structural transitions.}  of width $\sigma=1.7K$.

\begin{figure}[th]
\includegraphics[width=\columnwidth]{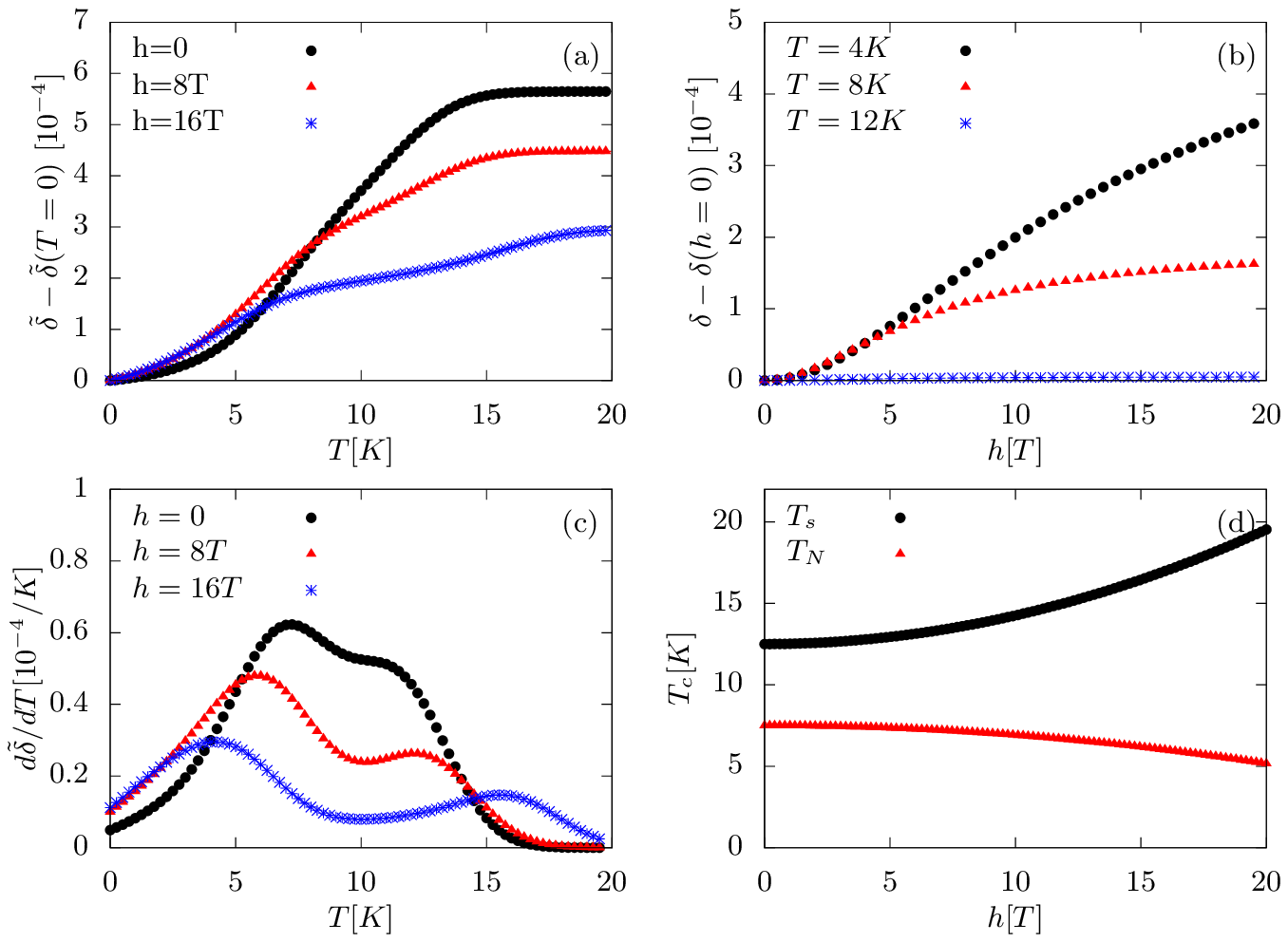}
	\caption{ Gaussian convoluted order parameter $\tilde{\delta}$ (a) as a function of the temperature and (b) magnetic field. (c) Linear distortion parameter $d\tilde{\delta}/dT$. (d) Structural $T_s$ and magnetic $T_N$ transition temperatures as a function of the external magnetic field.}
\label{fig:convo}
\end{figure}

Figure \ref{fig:convo} presents the numerical results for $\tilde{\delta}$ where, to ease the comparison with the experimental data, the values of $\tilde{\delta}$ are shifted to make them equal to zero at $T=0$ in figure \ref{fig:convo}(a), and for $h=0$ in figure \ref{fig:convo}(b). 
The thermal expansion $ \tilde{\alpha}_L=d\tilde{\delta}/d T$  presents a broad asymmetric structure at zero field which splits as the magnetic field is increased into a low temperature peak associated with the N\'eel transition and a high temperature peak due to the structural transition.\footnote{It may seem surprising that the peak associated with the structural transition has a lower height than the one associated with the N\'eel transition. After the Gaussian convolution, however, the height of the peaks is determined by the area of $d\delta/dT$ near the transition and not by the original height of the peaks.} Due to the asymmetry of $d\delta/dT$ near the transitions, the peaks in $d\tilde{\delta}/d T$ are shifted to lower temperatures than in the raw data.

\section{Conclusions}
We developed a Landau theory to describe phenomenologically the thermal expansion, magnetostriction, magnetization and specific heat data of CeCo$_{0.85}$Fe$_{0.15}$Si polycrystalline samples. 
We find that the available experimental data is compatible with the presence of a structural transition at a temperature $T_s(B)$ and a magnetic transition at $T_N(B)<T_s(B)$. The system presents a strong magnetoelastic coupling which leads to an increase in the structural transition temperature with an increasing external magnetic field, and allows the observation of a signature of the magnetic transition in the thermal expansion data.
Additional experimental data would be necessary to determine the nature of the structural transition. In particular, it would be important to determine the type of magnetic order and whether it breaks the tetragonal symmetry, inducing a lattice distortion. 
A magnetoelastic analysis of additional samples with Fe concentrations $y<0.2$ would allow to determine whether the structural transition temperature $T_s$ follows the same doping behavior as the specific heat anomaly $T_A$ which could be associated with the onset of magnetic fluctuations. In the Fe pnictides, the magnetic fluctuations for temperatures  $T\gtrsim T_N$ drive a nematic transition which is concomitant with a structural transition \cite{kasahara2012electronic,qi2009global,capati2011nematic,fernandes2014drives}.
It would be particularly interesting to determine if CeCo$_{1-y}$Fe$_y$Si compounds have a magnetic and elastic behavior analogous to the one observed in the Fe pnictides. If this is the case, the structural transition would not break the symmetry in the sign of the order parameter $\delta$ (this can be obtained dropping the $c \delta^3$ term on (\ref{eq:elast}), which was small in our calculations). An anharmonic elastic coupling between atoms would lead to a change in the volume of the sample given by $\Delta L/L \propto \delta^2$ (see e.g. \cite{Winkelman1995}), but would not otherwise change the main conclusion of this work.





\section{acknowledgements}
PSC acknowledges insightful discussions with E. Jagla and J. Lorenzana. Financial support from SECTYP-UNCU 06/C489, PIP 0832 CONICET, PICT 201-0204, PIP 0576 CONICET.
 \bibliographystyle{iopart-num} 
 \bibliography{references}

\providecommand{\newblock}{}
\begin{thebibliography}{10}
\expandafter\ifx\csname url\endcsname\relax
  \def\url#1{{\tt #1}}\fi
\expandafter\ifx\csname urlprefix\endcsname\relax\def\urlprefix{URL }\fi
\providecommand{\eprint}[2][]{\url{#2}}

\bibitem{steglich2016foundations}
Steglich F and Wirth S 2016 {\em Reports on Progress in Physics\/} {\bf 79}
  084502

\bibitem{stewart1984heavy}
Stewart S~G 1984 {\em Reviews of Modern Physics\/} {\bf 56} 755

\bibitem{lohneysen2007fermi}
L{\"o}hneysen H~v, Rosch A, Vojta M and W{\"o}lfle P 2007 {\em Reviews of
  Modern Physics\/} {\bf 79} 1015

\bibitem{Petrovic2001}
Petrovic C, Pagliuso P~G, Hundley M~F, Movshovich R, Sarrao J~L, Thompson J~D,
  Fisk Z and Monthoux P 2001 {\em J.\ Phys. Condens. Matter\/} {\bf 13} L337

\bibitem{Movshovich2001}
Movshovich R, Jaime M, Thompson J~D, Petrovic C, Fisk Z, Pagliuso P~G and
  Sarrao J~L 2001 {\em Phys. Rev. Lett.\/} {\bf 86}(22) 5152--5155
  \urlprefix\url{http://link.aps.org/doi/10.1103/PhysRevLett.86.5152}

\bibitem{WALKER1997303}
Walker I, Grosche F, Freye D and Lonzarich G 1997 {\em Physica C:
  Superconductivity\/} {\bf 282-287} 303 -- 306 ISSN 0921-4534 materials and
  Mechanisms of Superconductivity High Temperature Superconductors V

\bibitem{Facio2015}
Facio J~I, Betancourth D, Pedrazzini P, Correa V~F, Vildosola V, Garc\'{\i}a
  D~J and Cornaglia P~S 2015 {\em Phys. Rev. B\/} {\bf 91}(1) 014409
  \urlprefix\url{https://link.aps.org/doi/10.1103/PhysRevB.91.014409}

\bibitem{0295-5075-53-3-354}
Petrovic C, Movshovich R, Jaime M, Pagliuso P~G, Hundley M~F, Sarrao J~L, Fisk
  Z and Thompson J~D 2001 {\em EPL (Europhysics Letters)\/} {\bf 53} 354
  \urlprefix\url{http://stacks.iop.org/0295-5075/53/i=3/a=354}

\bibitem{hulliger1979rare}
Hulliger F 1979 {\em Handbook on the physics and chemistry of rare earths\/}
  {\bf 4} 153--236

\bibitem{takeuchi1998magnetoelastic}
Takeuchi T, Haga Y, Iwasa K, Kohgi M and Suzuki T 1998 {\em Journal of
  magnetism and magnetic materials\/} {\bf 177} 463--464

\bibitem{rossat1980specific}
Rossat-Mignod J, Burlet P, Bartholin H, Vogt O and Lagnier R 1980 {\em Journal
  of Physics C: Solid State Physics\/} {\bf 13} 6381

\bibitem{siemann1979alternative}
Siemann R and Cooper B~R 1979 {\em Journal of Applied Physics\/} {\bf 50}
  1997--1999

\bibitem{siemann1979enhanced}
Siemann R and Cooper B~R 1979 {\em Physical Review B\/} {\bf 19} 2645

\bibitem{Correa2016}
Correa V~F, Betancourth D, Sereni J~G, Canales N~C and Geibel C 2016 {\em
  Journal of Physics: Condensed Matter\/} {\bf 28} 346003
  \urlprefix\url{http://stacks.iop.org/0953-8984/28/i=34/a=346003}

\bibitem{Sereni2014}
Sereni J~G, Berisso M~G, Betancourth D, Correa V~F, Canales N~C and Geibel C
  2014 {\em Phys. Rev. B\/} {\bf 89}(3) 035107
  \urlprefix\url{https://link.aps.org/doi/10.1103/PhysRevB.89.035107}

\bibitem{Sengupta2003}
Sengupta P, Sandvik A~W and Singh R~R~P 2003 {\em Phys. Rev. B\/} {\bf 68}(9)
  094423 \urlprefix\url{https://link.aps.org/doi/10.1103/PhysRevB.68.094423}

\bibitem{khomskii2010basic}
Khomskii D~I 2010 {\em Basic aspects of the quantum theory of solids: order and
  elementary excitations\/} (Cambridge University Press)

\bibitem{kasahara2012electronic}
Kasahara S, Shi H, Hashimoto K, Tonegawa S, Mizukami Y, Shibauchi T, Sugimoto
  K, Fukuda T, Terashima T, Nevidomskyy A~H {\em et~al.\/} 2012 {\em Nature\/}
  {\bf 486} 382--385

\bibitem{qi2009global}
Qi Y and Xu C 2009 {\em Physical Review B\/} {\bf 80} 094402

\bibitem{capati2011nematic}
Capati M, Grilli M and Lorenzana J 2011 {\em Physical Review B\/} {\bf 84}
  214520

\bibitem{fernandes2014drives}
Fernandes R, Chubukov A and Schmalian J 2014 {\em Nature physics\/} {\bf 10}
  97--104

\bibitem{Winkelman1995}
Winkelmann H, Gamper E, B\"uchner B, Braden M, Revcolevschi A and Dhalenne G
  1995 {\em Phys. Rev. B\/} {\bf 51}(18) 12884--12887
  \urlprefix\url{https://link.aps.org/doi/10.1103/PhysRevB.51.12884}

\end{thebibliography}





\end{document}